\documentclass[5p]{elsarticle}
\usepackage{lineno,hyperref}
\modulolinenumbers[5]
\usepackage{graphicx}   
\usepackage{latexsym}   
\usepackage{bm}

\begin{document}
\begin{frontmatter}


\title{Do lattice data constrain the vector interaction strength of QCD?}

\author{Jan Steinheimer and Stefan Schramm}

\address{Frankfurt Institute for Advanced Studies,
Johann Wolfgang Goethe Universit\"at,\\
Ruth-Moufang-Str. 1, 60438 Frankfurt am Main, Germany}
\ead{steinheimer@fias.uni-frankfurt.de}
\date{January 15, 2013}

\begin{abstract}
We show how repulsive interactions of deconfined quarks as well as confined hadrons have an influence on the baryon number susceptibilities and the curvature of the chiral pseudo critical line in effective models of QCD. We discuss implications and constraints for the vector interaction strength from comparisons to lattice QCD and comment on earlier constraints, extracted from the curvature of the transition line of QCD and compact star observables. Our results clearly point to a strong vector repulsion in the hadronic phase and near-zero repulsion in the deconfined phase.
\end{abstract}

\begin{keyword}
QCD vector interaction strength, baryon number susceptibilities
\end{keyword}

\end{frontmatter}

\section{Introduction}
Quantum Chromo Dynamics (QCD) at extreme conditions of temperature and/or density is a central topic of many experimental and theoretical investigations. Especially the transition from the hadronic to the quark-gluon phase is a key region for studies of hot matter in ultra-relativistic heavy-ion collisions as well as for the stability of hybrid stars consisting of hadrons and quarks.
Here, the role of the repulsive vector interaction in QCD has become a much discussed topic in recent literature. It is not only important for the general understanding of the strong interaction but also has concrete implications for effective models of QCD regarding the location of the hadron-quark transition \cite{Bratovic:2012qs,Contrera:2012wj,Lourenco:2012yv}. In heavy-ion phenomenology the vector repulsion is a possible explanation for the observed splitting in particle and anti-particle azimuthal momentum asymmetry \cite{Song:2012cd}, whereas in nuclear astrophysics the possibility of stars with quark matter core depends strongly on the existence of a quark vector repulsion \cite{Klahn:2011fb,Klahn:2013kga}. It is therefore of great interest to possibly constrain the strength of the hadronic and quark repulsive interaction from QCD itself.
In this work we propose that such an independent determination has been provided by the conserved charge susceptibilities evaluated with lattice QCD at $\mu_B=0$. These susceptibilities quantify the fluctuations of the conserved charges of QCD, in particular the net baryon number. These susceptibilities can provide constraints on the interaction of particles, as a repulsive interaction of baryons would have direct impact on the magnitude of the fluctuations. 
The purpose of this paper is to first reproduce the results from \cite{Kunihiro:1991qu}, though with a different model for the quark 
phase, but also advance on these conclusions in several ways. First we include a phenomenologically correct hadronic phase,
to achieve a satisfactory agreement with lattice QCD
thermodynamics in the separate phases. This improvement is 
necessary to make relevant statements on the importance of 
the different types of interactions. 
Secondly we also investigate the influence of the strength
of the vector repulsion on finite density properties of the 
QCD phase diagram, in particular on the curvature of the 
pseudo-critical line. This part is of essential interest
as there where several studies \cite{Bratovic:2012qs,Contrera:2012wj} which claimed
that the vector repulsion strength of QCD could be fixed solely
by its determination \bfseries at \mdseries the pseudo critical temperature at $\mu=0$,
with the curvature $\kappa$ of the pseudo critical line.
In the present paper we will show that this statement cannot be upheld.
In the following we will describe how we combine a hadronic and an effective quark model in order to construct an equation of state that gives a correct description of the two phases of QCD, the confined and deconfined phase connected by a smooth cross-over. We will then use this combined equation of state to investigate the sensitivity of the baryon number susceptibilities on possible hadronic and quark repulsive interactions. In particular we want to understand the role of the repulsive interaction in the two separate phases and relate our results to recent attempts to constrain the vector interaction strength \cite{Kunihiro:1991qu,Ferroni:2010xf,Steinheimer:2010sp}.

\section{The Combined Equation of State}
The total grand canonical potential of our model includes contributions from the hadrons ($\Omega_{had}$) the deconfined quarks ($\Omega_q$), as well as contributions from mean fields interacting with the hadrons and quarks ($V$) and the Polyakov loop potential ($U$).
\begin{equation}\label{eq1}
	\Omega_{tot}=\Omega_{had}+\Omega_q+V+U
\end{equation}

In the hadronic phase we use the parity doublet model for the baryon octet and add all hadronic resonances, with masses up to 2.2 GeV, in order to correctly describe the QCD thermodynamics below $T_c$. In the parity doublet model positive and negative parity states of the baryons are grouped in doublets. Their masses are generated by a coupling to the chiral field $\sigma$. The effective masses of the nucleon and its chiral partner then become: $m_{\pm}^*=\sqrt{(g_{\sigma}^{(1)}\sigma)^2+m_0^2}\pm g_{\sigma}^{(2)} \sigma$, where $g_{\sigma}^{(1)}$ and $g_{\sigma}^{(2)}$ are the scalar coupling parameters of the model. In the chirally restored phase, for vanishing $\sigma$, their masses are degenerate and identical to $m_0$ \cite{Dexheimer:2007tn,Steinheimer:2011ea}.
The hadronic contribution $\Omega_{had}$ therefore can be written as:
\begin{equation}
	\Omega_{had}=T \sum_{i \in H} \frac{\gamma_i}{(2 \pi)^3} \int d^3p \left[ \ln \left(1 \pm e^{-\frac{1}{T}\left[E^*_i\pm \mu^*_i\right]}\right)\right]
\end{equation}
where $\gamma_i$ is the hadronic degeneracy factor and $E^*_i=\sqrt{m_i^{*2}+p^2}$ is the single particle energy 
and $\mu_i^*=\mu_i-g_{V}^{B}\omega$ the effective chemical potential (see e.g. \cite{Asakawa:1989bq,Kitazawa:2002bc})
of the $i$'th hadronic species.
The scalar meson interaction driving the spontaneous breaking of chiral symmetry
can be written in terms of SU(3) invariants
$I_1 = Tr(\Sigma)  ~,~ I_2 = Tr(\Sigma^2) ~,~ I_3 = Tr (\Sigma^4) $. The full potential for the fields then becomes:
\begin{equation}
V = V_0 + \frac{1}{2} k_0 I_2 - k_1 I_2^2 - k_2 I_3 + \frac{1}{2} m_{\omega}^2 \omega^2
\end{equation}
where $V_0$ is fixed by demanding a vanishing potential in the vacuum, $\Sigma$ is the multiplet of the scalar mesons and $\omega$ the repulsive vector field. 
The free parameters $k_0=(368.8 \ \rm{MeV})^2$, $k_1=4.264$ and $k_2=-13.055$ and $g_{V}^{B}=5.563$ are fitted to describe nuclear ground state properties.
Within this approach we ensure a good description of well-known properties of nuclear matter properties by adjusting the baryonic attractive scalar and repulsive vector interaction strength.\\
To describe the transition from the confined hadronic phase to a deconfined quark phase we include explicitly the contributions
of the quarks and gluons in the thermodynamic potential, as discussed in detail in \cite{Steinheimer:2011ea}. This generates a smooth cross-over chiral and deconfinement transition at small chemical potential and high temperatures, and a first-order transition in the case of cold, dense systems like compact stars. 
The quarks and gluons are incorporated in a similar way as described in so-called Polyakov loop extended quark models \cite{Fukushima:2003fw,Fukushima:2008wg,Ratti:2005jh,Roessner:2006xn,Schaefer:2007pw}. In our implementation we add the thermal contribution of the quarks to the thermodynamic potential $\Omega_{tot}$:

\begin{equation}
	\Omega_{q}=-T \sum_{i\in Q}{\frac{\gamma_i}{(2 \pi)^3}\int{d^3k \ln\left(1+\Phi \exp{\frac{E_i^* - \mu_i^*}{T}}\right)}}
\end{equation}
where we sum over all three quark flavors. $\gamma_i$ is the corresponding degeneracy factor,
$\Phi$ is the Polyakov loop. For the anti quarks we have to use the conjugate of the Polyakov loop $\Phi^*$ and $-\mu_{\overline{q}}^*=\mu_{q}^*=\mu_{q}-g_{V}^{Q} \omega$. For the Combined EoS we keep the quark vector coupling to be $=0$, if not stated otherwise. The effective mass $m_{i}^{*} = m_0 + g^{i}_{ S \sigma} \sigma + g^{i}_{S \zeta} \zeta$ of the quarks (except a small bare mass term $m_0=55$ MeV) is generated through a coupling to the scalar fields $\sigma$ and $\zeta$, which correspond to the non-strange and strange scalar quark condensates, respectively. Here we chose coupling values of $g^{Q}_{S \sigma}=1.8$ and, following SU(3) relations, $g^{s}_{S \zeta}=\sqrt{2} \cdot g_{q S \sigma}$ in order to enable a smooth transition between the hadronic and quark part of the EoS.

The effective potential $U(\Phi,\Phi^*,T)$, which controls the dynamics of the Polyakov-loop, is also included in the thermodynamic potential. In our approach we adopt the ansatz proposed in \cite{Ratti:2005jh}:
\begin{eqnarray}
	U&=&-\frac12 a(T)\Phi\Phi^*\nonumber\\
	&+&b(T)ln[1-6\Phi\Phi^*+4(\Phi^3\Phi^{*3})-3(\Phi\Phi^*)^2]
\end{eqnarray}
 with $a(T)=a_0 T^4+a_1 T_0 T^3+a_2 T_0^2 T^2$, $b(T)=b_3 T_0^3 T$. The values of the parameters $a_0=3.51$, $a_1=-8.2$, $a_2=14.8$, $b_3= -1.75$ and $T_0= 156$ MeV where adjusted to get a reasonable description of the interaction measure from lattice QCD \cite{Borsanyi:2013bia} and have the correct asymptotic value for free massless gluons.
 
\begin{figure}[t]	
\center \includegraphics[width=0.5\textwidth]{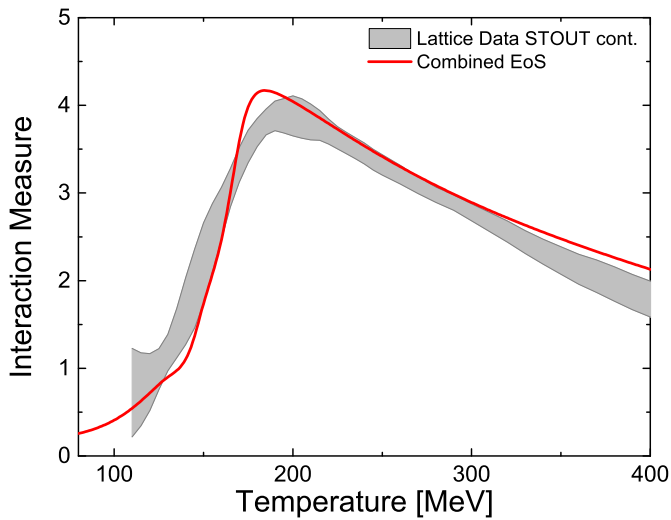}	
\caption{Interaction measure, $(\epsilon -3p)/T^4$, for our combines equation of state (red solid line) compared to
continuum extrapolated lattice QCD results \cite{Borsanyi:2013bia} (grey band). 
}\label{f1}
\end{figure}

To suppress hadrons when deconfinement is realized we adopt an ansatz introduced in \cite{Rischke:1991ke} and used in \cite{Steinheimer:2010ib}, where we introduced an excluded volume for the hadrons (and not the quarks), which very effectively removes the hadrons once the free quarks give a significant contribution to the pressure. In this approach the excluded volume $v_i$, of particle species $i$, enters in the total volume as $V=V'+\sum{v_i \cdot N_i}$, where $V'$ is the volume not occupied and $N_i$ the number of particles i in a volume. From this simple relations follows that one can also rewrite the
chemical potential $\mu_i$ of a particle species as $\tilde{\mu_i}=\mu_i -v_i P$ 
with $P$ the total pressure of all particle species.
This volume parameter $v_i$ is usually set to be a fixed value, but in this work we have chosen to make it explicitly dependent on the temperature. Such a dependence has the advantage that we can better match the combined equation of state to available lattice data. 
To make sure all densities (energy density and entropy) are thermodynamically consistent the usual densities $e$,$\rho$,$s$ have to be multiplied with a volume correction factor $f$, which is the ratio of the total volume $V$ 
and the reduced volume $V'$, not being occupied, yielding the corrected quantities ($\widetilde{e_i}$, $\widetilde{\rho_i}$ and $\widetilde{s_i}$).
The temperature dependence of the volume parameter in this work is:
\begin{equation}
	v_i(T)=v_o \cdot f(T)
\end{equation}
with a simple sigmoid function $f(T)=	1/[1+ \exp(-(T-\tau)/\delta \tau)]$, 
where $\tau=156$ MeV corresponds to the transition temperature and $\delta \tau=8$ MeV is the width of the transition. The asymptotic excluded volume $v_o$ corresponds to a hadronic radius of $r = 0.84\,$fm. Such a dependence leads to a negligible correction at low temperature, as suggested from lattice results, and a strong suppression of hadrons in the deconfined phase. 
The physical interpretation of such a large volume can be understood as an expected significant phase space broadening of the hadronic states around $T=\tau$.
Note that due to the temperature dependent excluded volume the combined equation 
of state violates, to a certain degree,
thermodynamical consistency, but only in the close vicinity of the matching temperature.\\
By construction of the Polyakov Loop potential, one usually observes an appearance of free quarks even at temperatures considerably lower than $T_{PC}$. Even though their contribution to the thermodynamic quantities is very small compared to the hadronic contribution, we suppress the quarks in the confined phase in order to study the separate impact of hadrons and quarks on the susceptibilities and phase transition more clearly. To achieve this we also introduce an explicitly temperature-dependent mass term $\delta m_q^0$, which is added to the quark mass $m_q^*$:
  
\begin{equation}
	\delta m_q^0(T)=m^0 \cdot [1-f(T)]
\end{equation}
with corresponding values of $m^0 = 400$ MeV, $\tau = 130$ MeV and $\delta \tau = 3 $MeV.
Such a dependence essentially suppresses any contribution of the quarks below a temperature of $T \approx 130$ MeV.
The resulting interaction measure $(\epsilon -3p)/T^4$ is shown in figure \ref{f1} and compared with available lattice data. By adjusting the above mentioned parameters we obtain a very good description of the Interaction measure.

To isolate the contributions of the deconfined quarks we will also calculate the so called 'Quark EoS' which is essentially defined as the combined model, excluding the hadronic contribution to the grand canonical potential $\Omega_{tot}-\Omega_{had}$, but including all the potential terms. 

\section{Results}
\begin{figure}[t]	
\center \includegraphics[width=0.5\textwidth]{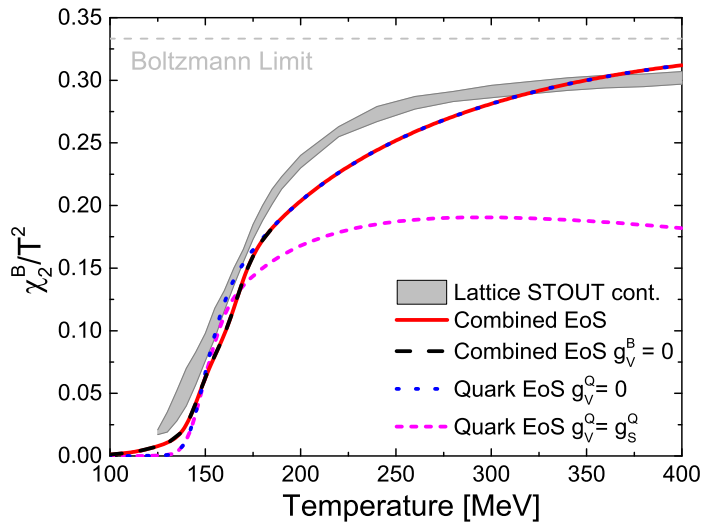}	
\caption{Second order baryon number susceptibility from our combined equation of state (red solid line) compared to
continuum extrapolated lattice QCD results \cite{Borsanyi:2011sw} (grey band). We also show the combined EoS with zero hadron vector interaction strength (black dashed line), and the quark part of the EoS with finite vector coupling (magenta short dashed line).
}\label{f2}
\end{figure}

Because lattice QCD suffers from the sign problem it is difficult to compute the QCD phase structure and thermodynamics at non-zero baryochemical potential $\mu_B$. It is however possible to infer information on the thermodynamics of QCD at small values of $\mu_B/T$ through a Taylor expansion of lattice results at $\mu_B=0$ in terms of the chemical potential \cite{Allton:2002zi}. In the Taylor expansion of the pressure $p=-\Omega$, the coefficients $c^B_n$, which can be related to the baryon number susceptibilities $\chi_n^{B}$, follow from:
\begin{equation}
 \chi_n^{B}/T^2 = n! c^B_n(T)=\frac{\partial^n(p(T,\mu_B)/T^4)}{\partial(\mu_B/T)^n}
\end{equation}
for $\mu_B=\mu_S=\mu_Q=0$.\\
As $p(T,\mu_B)$ also depends on the value of the vector field $\omega(T,\mu_B)$ explicitly one can easily see that
the susceptibilities have contributions which depend on the derivatives of this field $\partial^n \omega(T,\mu_B)/(\partial \mu_B)^n \ne 0 $. It is now interesting and instructive to investigate how large these contributions are and if one can use them to constrain
$\omega(T,\mu_B)$ and subsequently $g_V$.
In figure \ref{f2} we show our results for the baryon number susceptibility as a function of the temperature at $\mu_B=0$. As for the interaction measure we again obtain a good description of the lattice data, over the whole temperature range, from our combined EoS. The most interesting feature of this figure is the strong dependence of the second-order baryon number susceptibility on the value of the free quark repulsive interaction $g_V^Q$. Below the pseudo critical temperature $T_{PC}$ we observe hardly any change in $\chi_2^{B}$ due to the repulsive hadronic interaction strength $g_V^B$, but a strong decrease in $\chi_2^{B}$ above $T_{PC}$ due to a quark repulsive coupling $g_V^Q$.

\begin{figure}[t]	
\center \includegraphics[width=0.5\textwidth]{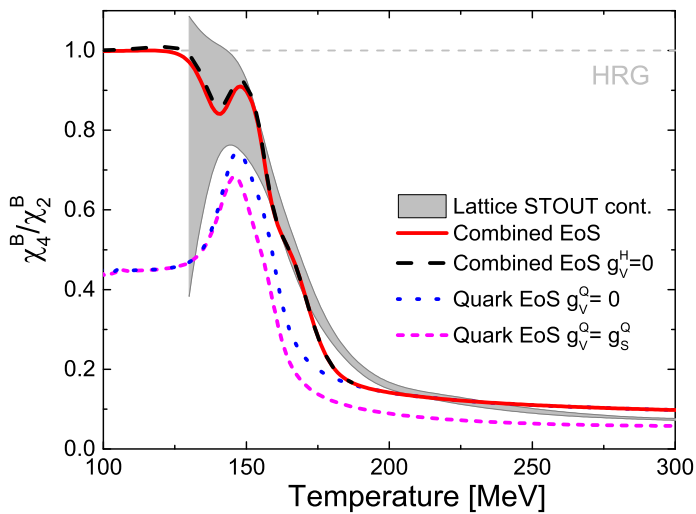}	
\caption{Forth order baryon number susceptibility from our combined equation of state (red solid line) compared to
continuum extrapolated lattice QCD results \cite{Bellwied:2013cta} (grey band). We also show the combined EoS with zero hadron vector interaction strength (black dashed line), and the quark part of the EoS 
where quarks have a finite vector coupling (magenta short dashed line).}\label{f3}
\end{figure}

A similar picture can be drawn from figure \ref{f3}, where the ratio of the fourth order over the second order baryon number susceptibilities is shown. Below $T_{PC}$, in the hadronic phase, we observe only a very small dependence of the susceptibility on the hadronic repulsive interaction strength, even though the dependence appears somewhat stronger for $\chi_4^{B}/\chi_2^{B}$ than for $\chi_2^{B}$. Such a weak dependence of the susceptibilities is understandable when we recall which hadronic degrees of freedom contribute to the susceptibilities. These are, in the case of the baryon number, dominantly nucleons and heavier baryons. Because these hadrons have a large mass, their density, and therefore any repulsive force, will only be significant at very large chemical potentials, comparable to their mass. At such high chemicals potentials, the lower-order susceptibilities will not be the relevant contributors to a Taylor expansion, but rather higher-order terms. To understand the influence of the repulsive interactions of hadrons one therefore would have to evaluate susceptibilities of even higher order. Because the quarks have significantly smaller masses, once chiral symmetry is restored, they contribute strongly also to lower orders of the baryon number susceptibility which allows for a much stricter constraint on the quark repulsive interaction strength. 
A similar behavior was found already in the NJL model by Kunihiro where a formula was derived (eq. 3.12 in \cite{Kunihiro:1991qu}) which showed
the effect of the mass reduction due to the chiral restoration and the vector coupling was considered as an additional mechanism to account for the lattice data in \cite{Kunihiro:1991qu}. Note that because \cite{Kunihiro:1991qu} did not include a transition from hadrons to quarks it was not possible to fully disentangle the different contributions, of the phases, to the susceptibilities.

\begin{figure}[t]	
\center \includegraphics[width=0.5\textwidth]{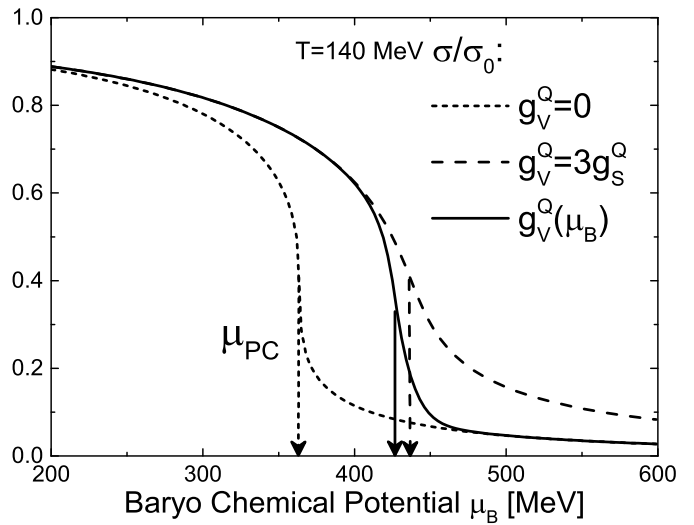}	
\caption{The normalized chiral condensate, order parameter of the chiral phase transition, as a function of chemical potential at fixed temperature $T=140$ MeV. We compare different scenarios of the repulsive quark vector interaction strength. The vertical lines indicate the pseudo critical chemical potentials, defined from the maximum of the chiral susceptibility. 
}\label{f4}
\end{figure}		

In the deconfined phase, for $T>T_{PC}$ we again observe a significant deviation of the calculated values of $\chi_4^{B}/\chi_2^{B}$ from lattice results, whenever we assume a repulsive interaction between the free quarks. As the ratio decreases by a factor of $1/2$ above $T_{PC}$ we can conclude that the relative contribution of the repulsive interaction to $\chi_4^{B}$ is considerably larger than for $\chi_2^{B}$.

In several publications (e.g. \cite{Bratovic:2012qs,Contrera:2012wj,Lourenco:2012yv,Klahn:2011fb,Klahn:2013kga}) it has been argued that a non-zero repulsive quark vector coupling is required by constraining the interior of compact stars and the phase structure of constituent quark mean field models. For example one can show that the newly measured maximum mass of compact stars of 2 solar masses can only be accommodated for, if one either assumes a very small or no quark content for these stars, or introduces a quite sizable repulsive interaction for quarks. For isospin symmetric matter, it has been shown that constituent-quark based models like the PNJL and PQM model can only accommodate for a ''nuclear ground'' state (in these models this is approximated by a constituent-quark saturated state) if the quarks have a finite vector interaction strength. We therefore have to ask the question if our results are in contradiction to these earlier attempts on constraining the quark repulsive interaction.

\begin{table}[h]
	\centering
		\begin{tabular}{|c|c|c|c|c|c|c|}
		\hline 
	EoS: & \multicolumn{2}{|c|}{I}   & \multicolumn{2}{|c|}{II} & \multicolumn{2}{|c|}{III} \\ \hline	
 $g_{V}^{Q}=$	&	0     & $3 g_{S}^{Q}$ &  0  & $3 g_{S}^{Q}$ &  0  & 0 \\ \hline
 $g_{V}^{N}=$	&	5.563 & 5.563         &  0  &            0  &  0  & 5.563 \\ \hline 
 $\kappa=$    &	0.143 & 0.1           &0.139& 0.0463        &0.25 & 0.175 \\ \hline
		\end{tabular}
	\caption[kappa]{Values of the curvature of the chiral pseudo critical line from different EoS.
	 I: The Combined EoS,
	 II: The Quark EoS and
	 III: From the Hadronic EoS only.}\label{kappa}
\end{table}

To give a possible answer, we investigate the curvature of the chiral pseudo critical line for different realizations of our model.
In \cite{Kaczmarek:2011zz} the curvature
of the pseudo critical line is estimated within lattice QCD
using the light quark chiral condensate.
Here the chiral condensate $\left\langle \overline{\Psi}\Psi\right\rangle(T)$
is expanded in terms of the chemical potential as 
 $\left\langle \overline{\Psi}\Psi\right\rangle(T,\mu)=
 \left\langle \overline{\Psi}\Psi\right\rangle(T,0)+\chi_{m,q}/2\mu^2T$
where $\chi_{m,q}=\partial\chi_q/\partial m_l$ and $\chi_q$ is the light quark
number susceptibility, taken only to second order and $m_l$ the light quark mass.
To find the curvature $\kappa$ one has to find the maximum
in $\partial \left\langle \overline{\Psi}\Psi\right\rangle(T,\mu)/ \partial T$
which gives $T_{PC}(\mu)$ and is related to the curvature $\kappa \propto \partial^2 T_{PC}/\partial \mu^2$.
Note that only the second order susceptibilities where included in
the lattice study.
To second order the curvature $\kappa$ therefore depends on
the light quark number susceptibility, but also its derivatives
with respect to $T$ and $\mu$ at $\mu=0$ and $T=T_{PC}$. 
Therefore it is not simply a function of the second order 
susceptibility, but does depend on the (4th order) susceptibility
near $T_{PC}$ at $\mu=0$.
Defining the shifting $T_{PC}$ with chemical potential requires higher-order susceptibilities.
Overall, as we showed, the 
vector interaction strength may vary rapidly around that 
point and therefore the determination of $g_V$
from $\kappa$ is not very reliable as also the repulsive interaction strength may change rapidly around $T_c$\\
We calculated the curvature of the pseudo critical line, according to the chiral transition, from our combined model, and the quark and hadronic phases separately
(i.e. leaving out the hadronic or quark contribution to eq.(\ref{eq1})). The results for the curvature $\kappa$ are presented in Table \ref{kappa}, where we show 
the two cases where the vector repulsion strength of hadrons and quarks is either set to zero everywhere or is a fixed finite value.
For all equations of state we observe that the transition for finite vector couplings is moved to larger chemical potentials, i.e. the curvature of the transition is decreased, as required by \cite{Bratovic:2012qs,Contrera:2012wj,Lourenco:2012yv}. 

To isolate the effect of the appearance of the quarks and their vector interaction, we introduce a chemical dependent quark vector coupling strength $g_V^Q(\mu_B)$ in a schematic way, which we can define such that it disappears at the pseudo critical line.
For our result in figure \ref{f4} this would correspond to:
\begin{equation}
	g_{V}^Q(\mu_B)= g_{V}^Q(\mu_B=0) \cdot (1+\exp(\mu_B - \mu_B^{PC})/\delta_{\mu})^{-1}
\end{equation}
where $\mu_B^{PC}$ is the pseudo-critical chemical potential for a constant $g_{V}^Q=g_{q\sigma}$, and $\delta_{\mu}=10$ MeV. The resulting curve of the normalized chiral condensate is presented as solid line in figure \ref{f4} where we show the behavior of the normalized chiral condensate $\sigma/\sigma_0$, from our combined equation of state, as a function of baryochemical potential, for an arbitrarily fixed temperature $T=140$ MeV relatively close to, but below, $T_c$. Here the short dashed line represents the result for vanishing quark vector coupling. We also show the same curve with finite quark vector coupling as dashed line. One can clearly see that we obtain a larger pseudo-critical chemical potential, comparable to that with large quark vector repulsion, even though the repulsive strength is vanishing in the 'deconfined' phase. The shift in $T_{PC}$ is therefore determined by the behavior of the interacting matter below the pseudo-critical temperature and the appearance of the free quarks.\\
  
The apparent contradiction of \cite{Bratovic:2012qs,Contrera:2012wj,Lourenco:2012yv} with our result, stating that there should be no quark repulsive interaction, can be explained in a very simple way. In PNJL and PQM type models, only one type of vector coupling strength exists, which is the same for light ``unconfined'' quarks above the transition as well as possible three-quark states that appear in the confined phase. In reality of course this might not hold true. We know that there should be no quarks in the confined phase and we also have strong indications for a considerable repulsive hadronic interaction.  
In other words the large repulsive vector interaction, required in the PQM and PNJL models might be simply a result of the unsatisfactory description of the confined, or hadronic, phase. 

\section{Summary}
We have shown that lattice results for the baryon number susceptibilities can be used, even to lowest order, to constrain the repulsive vector interaction strength of quarks in the deconfined phase. We confirm \cite{Kunihiro:1991qu} that only a nearly vanishing strength is supported by lattice QCD data. Even a small vector coupling would lead to a systematic deviation of the baryon number susceptibilities, i.e. a maximum as function of temperature at $\mu_B=0$. Such a behavior is not observed, even when susceptibilities are calculated to very high temperatures on the lattice \cite{Bazavov:2013uja} and in perturbative QCD \cite{Haque:2013sja}. Concerning the repulsive hadronic interaction we find that, due to the large mass of the baryonic hadrons, the lowest order susceptibilities show only a very weak dependence and are not useful to constrain the hadronic repulsive interactions.
We also show that earlier constraints on the quark repulsive coupling, using the curvature of the transition line and compact star masses are not as strict as they claim to be. Both can also be accommodated taking into account a more realistic hadronic phase and a strong change of the vector coupling strength at the deconfinement transition (appearance of free quarks). As an example, we show that the curvature of the transition is sensitive to the repulsive interaction in the confined phase and not necessarily in the deconfined phase. Also compact stars with large masses can be accommodated for with stiff hadronic equations of state, due to hadronic repulsive forces.\\
Concluding, we believe that the results shown represent a considerable step forward in the understanding of the interactions of deconfined quarks. Furthermore they present a strict set of  constraints for effective models of QCD.

\section{Acknowledgments}
We wish to acknowledge stimulating discussion with Volker Koch. This work was supported by GSI and the Hessian initiative for excellence (LOEWE) through the Helmholz International Center for FAIR (HIC for FAIR).



			\end{document}